
%
%
%
%

\input harvmac
\def \x{\xi}

\def \s{\sigma}
\def \b{\beta}
\def \a{\alpha}
\def \g{\gamma}
\def \d{\delta}
\def \e{\epsilon}
\def \l{\lambda}

\def \ph{\phi}

\def \p{\pi}

\def \G{\Gamma}

\def \m{\mu}
\def \n{\nu}

\def \O{\Omega}

\def \r{\rho}

\def\apny{Ann.\ Phys.\ (New York)\ }
\def\cmp{Comm.\ Math.\ Phys.\ }

\def\ijmpa{{Int.\ J.\ Mod.\ Phys.\ }{\bf A}}

\def\mpla{{Mod.\ Phys.\ Lett.\ }{\bf A}}

\def\npb{{Nucl.\ Phys.\ }{\bf B}}

\def\plb{{Phys.\ Lett.\ }{\bf B}}

\def\prl{Phys.\ Rev.\ Lett.\ }

\def\lmp{Lett.\ Math.\ Phys.\ }

\def \pri{\prime}

\def \del {\partial}

\def \cG {{\cal G}}

\nopagenumbers
\line {\hfil LTH 317}
\vskip .5in
\centerline{\titlefont The Exact Tachyon Beta-Function}
\vskip 5pt
\centerline {\titlefont for the }
\vskip 5pt
\centerline {\titlefont Wess-Zumino-Witten
Model}
\vskip 10pt
\centerline{I. Jack and D. R. T. Jones}
\bigskip
\centerline{\it DAMTP, University of Liverpool, Liverpool L69 3BX, U.K.}
\vskip .3in
We derive an exact expression for the tachyon $\b$-function for the
Wess-Zumino-Witten model. We check our result up to three loops
by calculating the three-loop
tachyon $\b$-function for a general non-linear $\s$-model with torsion,
and then specialising to the case of the WZW model.
\Date{October 1993}

\newsec{Introduction}
The Wess-Zumino-Witten (WZW) model\ref\ew{E. Witten, \cmp92 (1984) 483.}\
 is a particularly interesting
example of a conformal field theory. Indeed it is currently believed
that all rational conformal field theories may be derived from the WZW
model by the Goddard-Kent-Olive (GKO) construction\ref\GKO{P. Goddard, A. Kent
and D. Olive, \plb152 (1985) 88.}
 (or equivalently
by gauging\ref\Gaug{K. Gawedzki and A. Kupiainen, \plb215 (1988) 119;
\npb320 (1989) 625\semi
D. Karabali and H. J. Schnitzer, \npb329 (1990) 649.}).
The WZW model on a Lie group manifold $\cG$ is parametrised
by the level (which is constrained to be an integer for a compact
$\cG$). The properties which characterise the conformal field
theory--the central charge and the conformal dimensions of the primary
fields--have exact non-perturbative expressions in terms of the level
 and the Casimirs for the Lie algebra of $\cG$\ref\kz{V. G. Knizhnik and
A. B.Zamolodchikov, \npb247 (1984) 83.}.

In general, we may consider perturbing the WZW model by adding a
potential term to the action. This potential term is usually taken to be
a primary field of the WZW model; however we would like to consider the
case where this restriction is not applied. Our motivation for this
comes from non-abelian Toda field theories\ref\ls{A. N. Leznov and
M. V. Saveliev, \lmp6 (1982) 505; \cmp89 (1983) 59.}, which have recently
been receiving some attention\ref\natta{M. V. Saveliev,
\mpla5 (1990) 2223.}
\ref\gs{J.-L. Gervais and M. V.
Saveliev, \plb286 (1992) 271\semi
F. Delduc, J.-L. Gervais and M. V. Saveliev, \plb292 (1992) 295\semi
L. O'Raifeartaigh and A. Wipf, \plb251 (1990) 361\semi
L. O'Raifeartaigh, P. Ruelle, I. Tsutsui and A. Wipf, \cmp143 (1992)
333\semi
L. Feh\'er,
L. O'Raifeartaigh, P. Ruelle, I. Tsutsui and A. Wipf,
\apny213 (1992) 1\semi
G. Papadopoulos and B. Spence, ``The Space of Solutions of Toda Field
Theory'', preprint UM-P-93/38, KCL-93-7, hep-th/9306088.}.
The action for these models
consists precisely of a WZW model coupled to a potential term. In a
recent paper\ref\natt{I. Jack and D. R. T. Jones,
``Quantum non-abelian Toda field theories'',
Liverpool preprint LTH 315.}\
 we discussed the conformal properties of the non-abelian
Toda field theory at the quantum level. The conformal behaviour of the
potential term is described by the tachyon $\b$-function (in terminology
derived from string theory) and therefore we needed an exact expression
for this $\b$-function. Our aim in this paper is to derive this. In the
particular case where the potential is simply $\tr(g)$, where $g\in\cG$,
the tachyon $\b$-function is simply related to the conformal dimension
of $\tr(g)$, and we shall exploit this to deduce the general form of the
tachyon $\b$-function. We shall then check this result by performing an
explicit perturbative calculation up to three-loop order. We first
obtain the result for a general non-linear $\s$-model, and then specialise
to the case of the WZW model. In the WZW case, our perturbative
calculation is similar to a three-loop calculation of the conformal
dimension of $\tr(g)$ carried out some time ago\ref\bos{M. Bos, \apny181 (1988)
177}. However, we use a
different prescription\ref\ht{
C. M. Hull and P. K. Townsend, \plb191 (1987) 115\semi
R. R. Metsaev and A. A. Tseytlin, \plb191 (1987) 354;\npb293 (1987) 385\semi
D. Zanon, \plb191 (1987) 363\semi
D. R. T. Jones, \plb192 (1987) 391.}
\ref\hugh{H. Osborn, \apny200 (1990) 1.}
 for continuing the two-dimensional alternating
symbol to $d$ dimensions in the context of dimensional regularisation.
This prescription has the conceptual advantage of avoiding the need for
introducing extra evanescent terms into the action, and also preserves
explicit $O(d)$ covariance. It now appears to have been accepted as the
standard prescription for calculations of this type.
\newsec{Exact result for tachyon $\b$-function}
 In this Section we derive our principal result, namely the exact
tachyon $\b$-function for the WZW model. We first write down the action for
the Wess-Zumino-Witten (WZW) model defined on a group
manifold $M_{\cG}$:
\eqn\eaa{
kS_{WZW}(g)=-{k\over{8\p}}\int_S d^2x \tr(g^{-1}\del_{\m} gg^{-1}\del^{\m}g)
+{ik\over{12\p}}\int_B d^3x \e^{\m\n\r}\tr(g^{-1}\del_{\m} gg^{-1}\del_{\n}g
g^{-1}\del_{\r}g)  }
where $g$ is a group element in the defining representation of $\cG$,
whose generators satisfy
\eqn\eab{
[T_a,T_b]=if_{abc}T_c.}
$B$ is a 3-dimensional ball whose
surface is the two-dimensional worldsheet $S$. We assume that the group
generators satisfy $\tr(T_aT_b)=\d_{ab}$.
We are using here the conventions of Ref. \ref\go{P. Goddard and D. Olive,
\ijmpa1 (1986) 303.}. The level $x$ is defined in terms of $k$ by
$x={2k\over{\psi^2}}$, where $\psi$ is the highest root in
$\cG$.
The central charge is given by
\eqn\eac{
c={k{\rm dim}\cG\over{k+{1\over2}c^{\cG}}},  }
and $g$ is a primary field with conformal dimensions given by\kz
\eqn\ead{
h=\bar h={c^R\over{k+{1\over2}c^{\cG}}},  }
where $c^{\cG}$ and $c^R$ are the eigenvalues of the quadratic Casimir in
the adjoint and defining representations respectively, so that
\eqn\eaf{
f_{acd}f_{bcd}=c^{\cG}\d_{ab}, \qquad T_aT_a=c^R1  .}
The WZW model is a particular example of a two-dimensional non-linear
$\s$-model, whose action is given in general by
\eqn\eag{
S(\ph)={\l\over{8\pi}}\int d^2x\sqrt{\g}
\{\g^{\m\n}G_{ij}(\ph)\del_{\m}\ph^i\del^{\n}\ph^j
+\e^{\m\n}B_{ij}(\ph)\del_{\m}\ph^i\del_{\n}\ph^j
 +{1\over{\l}}D(\ph)R^{(2)}+V(\ph)\}  }
where $\g_{\m\n}$ is the two-dimensional metric, $\g$ is its determinant and
$\e^{\m\n}$ is the two-dimensional alternating symbol.
$\{\ph^i\}$ represent co-ordinates on some target manifold with
metric $G_{ij}$ and antisymmetric tensor field $B_{ij}$ defined on it,
$D$ is the dilaton field coupling to the two-dimensional scalar curvature
$R^{(2)}$, and $V$ is the tachyon field. (The terminology derives from string
theory; our conventions are equivalent to taking $\a^{\pri}={2\over{\l}}$
in Ref. \hugh, where $\a^{\pri}$ is the string coupling.)
The conformal invariance conditions for the $\s$-model Eq. \eag\ may
be written\hugh
\eqna\eh
{$$\eqalignno{
B^G_{ij}&\equiv\b_{ij}^G+{2\over{\l}}\nabla_i\del_jD+2\del_{(i}W_{j)}=0
&\eh a\cr
B_{ij}^B&\equiv\b_{ij}^B+{2\over{\l}}H^k{}_{ij}\del_kD+2H^k{}_{ij}W_k=0
&\eh b\cr
B^V&\equiv\b^V-2V+{1\over{\l}}\del^kD\del_kV+W^k\del_kV=0&\eh c\cr}$$   }
where $\b^G_{ij}$, $\b^B_{ij}$ and $\b^V$ are the standard renormalisation
group $\b$-functions for $G_{ij}$, $B_{ij}$ and $V$, and $H_{ijk}$ is the
torsion, defined by $H_{ijk}=3\nabla_{[i}B_{jk]}$. $W_i$ is a vector field
which can be determined perturbatively within a given renormalisation
scheme. We shall be using
a renormalisation scheme in which $W_i$ vanishes for the WZW model.

When $B^G_{ij}$ and $B^B_{ij}$ both vanish, the quantity $B^D$ given by
\eqn\eai{
B^D\equiv\b^D+{1\over{\l}}\del^kD\del_kD+W^k\del_kD, }
where $\b^D$ is the dilaton $\b$-function, becomes constant
\ref\CP{G. Curci and G. Paffuti, \npb286 (1987) 399.}\hugh\ and is then
related to the central charge $c$ for the conformal field theory by
\eqn\eaj{
c=3B^D. }
The tachyon $\b$-function is of the form
\eqn\eal  {
\b^V=\O V  }
where $\O$ is a differential operator, in general of arbitrary order.
We may write $\O$ in the form
\eqn\eam  {
\O=\sum X^{(n)k_1\ldots k_n}\nabla_{k_1}\ldots\nabla_{k_n} }
where $X^{(n)k_1\ldots k_n}$ is an $n$th rank tensor constructed from
the Riemann tensor, the torsion and their derivatives contracted together.

In the case of the WZW model Eq. \eaa, the target manifold is the group
$M_{\cG}$.
There is no tachyon field $V$, and the
metric $G_{ij}$ and $B_{ij}$ may be read off by comparing Eqs. \eaa\ and
\eag. We have\ref\braat{E. Braaten, T. L. Curtright and C. Zachos,
\npb260 (1985) 630.}
\eqn\eak{
G_{ij}=e_{ai}e_{aj}, \qquad \l=k, }
where the vielbein $e_{ai}$ is defined by
\eqn\eaka{
ig^{-1}\del_ig=e_{ia}T_a }
and satisfies
\eqn\eakab{
e_{a}{}^ie_{bi}=\d_{ab} .}
In fact $B_{ij}$ can only be defined locally, but we have a globally defined
expression for $H_{ijk}$\braat,
\eqn\eakb{
H_{ijk}={1\over2}f_{abc}e_{ai}e_{bj}e_{ck}. }
We also have
\eqn\eakc{
\nabla_ie_{aj}=f_{abc}e_{bi}e_{cj}  }
from which it follows that the Riemann tensor is given by\braat
\eqn\eakd{
R_{ijkl}={1\over4}f_{abe}f_{cde}e_{ai}e_{bj}e_{ck}e_{dl} }
and also that
\eqn\eake{
\nabla_lH_{ijk}=0  .}

We will denote quantities
evaluated for the particular case of the WZW model
by subscripts $WZW$. $\O_{WZW}$ has an
expansion similar to Eq. \eam, but in terms of $X_{WZW}^{(n)k_1\ldots k_n}$
obtained from $X^{(n)k_1\ldots k_n}$ by
substituting the expressions in Eqs. \eakb\ and \eakd\
 for the Riemann tensor and
torsion. $X_{WZW}^{(n)k_1\ldots k_n}$ is constructed purely from the structure
constants and the vielbeins.

If $V$ is a primary field, it will be an eigenfunction of $\O_{WZW}$ whose
eigenvalue is the conformal dimension of $V$. In particular, for $V=\tr(g)$,
we must have
\eqn\ean  {
\O_{WZW}\tr(g)={c^R\over{k+{1\over2}c^{\cG}}}\tr(g). }
This information is sufficient to determine $\O_{WZW}$. We can assume without
loss of generality that $X_{WZW}^{(n)k_1\ldots k_n}$ is symmetric in
$k_1 \ldots k_n$ (indeed, $X_{WZW}^{(n)k_1\ldots k_n}$ is naturally given in
this form by the explicit calculation--see later). So we have, using Eqs.
\eaka\
and \eakc,
\eqn\eao{
X_{WZW}^{(n)k_1\ldots k_n}\nabla_{k_1}\ldots\nabla_{k_n}\tr(g)=
i^nX_{WZW}^{(n)k_1\ldots k_n}\tr(gT_{a_1}\ldots T_{a_n})e_{a_1k_1}\ldots
e_{a_nk_n}
,  }
since the structure constant terms in Eq.~\eakc\ vanish by symmetry here.
Eq. \ean\ can be expanded in powers of $1\over k$ (or
equivalently in powers of $c^{\cG}$), and this expansion must
correspond to the perturbation expansion for $\O$.
Hence we see that
$X_{WZW}^{(n)k_1\ldots k_n}T_{a_1}\ldots T_{a_n}e_{a_1k_1}\ldots e_{a_nk_n}$
must reduce to a function of $c^R$ and $c^{\cG}$ linear in $c^R$.
The only way in which this can happen is if
\eqn\eap{
X_{WZW}^{(n)k_1\ldots k_n}=0,\qquad i\ne2  }
\eqn\eaq{
X_{WZW}^{(2)k_1k_2}T_{a_1}T_{a_2}e_{a_1k_1}e_{a_2k_2}
=-{c^R\over{k+{1\over2}c^{\cG}}}.}
We then must have
\eqn\ear{
X_{WZW}^{(2)k_1k_2}e_{a_1k_1}e_{a_2k_2}=-{c^R\over{k+{1\over2}c^{\cG}}}
\d_{a_1a_2}, }
and hence, from Eq. \eak,
\eqn\eas{
X_{WZW}^{(2)k_1k_2}=-{c^R\over{k+{1\over2}c^{\cG}}}g^{k_1k_2}.}
So we finally have the desired result
\eqn\eat{
\O_{WZW}=-{c^R\over{k+{1\over2}c^{\cG}}}\nabla^2, }
and hence
\eqn\eau{
\b^V_{WZW}=-{c^R\over{k+{1\over2}c^{\cG}}}\nabla^2V.  }
\newsec{Three-loop perturbative calculation}
In this Section we go some way towards corroborating our exact result for the
tachyon $\b$-function for the WZW model, derived in the previous Section,
by performing a perturbative calculation up to three-loop order. We
start by doing the computation for the general non-linear $\s$-model of Eq.
\eag, before specialising to the WZW model. It is most convenient to perform
these computations using dimensional regularisation; however, the crucial
issue which then arises for a $\s$-model of the form Eq. \eag, one with a term
involving the antisymmetric tensor $B_{ij}$, is how to extend the
two-dimensional alternating symbol, $\e^{\m\n}$, away from two dimensions.
In two dimensions one has the relation
\eqn\eba{
\e^{\m\n}\e^{\r\s}=\g^{\m\s}\g^{\n\r}-\g^{\m\r}\g^{\n\s} .}
However, if one tries to apply this relation for $d\ne2$, one encounters
inconsistencies. One solution is to regard $\e^{\m\n}$ as strictly
two-dimensional even within dimensional regularisation\bos. The drawback of
this approach is that the tangent-space group is reduced from $O(d)$ to
$O(d-2)\times O(2)$, and as a consequence one is obliged to introduce
additional, ``evanescent'' couplings which were not present in the original
two-dimensional action. Physical results independent of these evanescent
couplings can be obtained, but only at the expense of additional
calculation. This is the method used in Ref. \bos\ to compute the anomalous
dimension of $\tr(g)$ up to three loops.

An alternative approach, which has been fairly widely used\ht\hugh, is to
abandon Eq. \eba\ away from two dimensions, but to assume the existence
of a tensor
$\e^{\m\n}$ in general $d$ dimensions with the property
\eqn\ebb{
\e^{\m\r}\e_{\r}{}^{\n}=-\g^{\m\n}   .}
We should stress that the $\g^{\m\n}$ which appears on the RHS of Eq. \ebb\
is the $d$-dimensional metric. Evanescent couplings are required in this case
also for the rigorous discussion of renormalisability\hugh; however,
the important
difference as compared to the previous prescription is that these evanescent
couplings completely decouple from physical results. Moreover it turns out
that a relation of the form Eq. \ebb\ is sufficient for perturbative
calculations.

The most convenient means for discussing the quantisation of the non-linear
$\s$-model is the use of the background field method\ref\afm{L.
Alvarez-Gaum\'e, D. Z. Freedman and S. Mukhi, \apny134 (1981) 85.}
. The field $\ph^i$ is
expanded around a classical background configuration as
\eqn\ebc{
\ph^i=\ph_0^i+\p^i  . }
However, the field $\p^i$ is not very convenient for use as the quantum
field variable, since it does not transform as a vector. It is customary to
write $\p^i$ in terms of the field $\x^i$, the tangent vector to the
geodesic linking $\ph_0^i$ to $\ph_0^i+\p^i$, and to use $\x^i$ as the
quantum field\afm\ref\muk{S. Mukhi, \npb264 (1986) 640.}
. This guarantees a manifestly covariant perturbation
expansion written in terms of tensor quantities such as the Riemann tensor,
the torsion and their covariant derivatives. The expansion of the action
for the non-linear $\s$-model in Eq. \eag\ in terms of $\x$ takes the
following form (setting the dilaton term, which will not concern us, to zero)
\afm\ref\fv{B. E. Fridling and A. E. M. van de Ven, \npb268 (1986) 719.}\muk:
\eqn\ebd{
S(\ph+\p)=S(\ph)+\sum_{i=1}S^{(i)}(\ph,\x) , }
where
\eqna\ebe{$$\eqalignno{
S^{(1)}&=\int d^2x(g_{ij}\del_{\m}\ph^iD^{\m}\x^j+\e^{\m\n}H_{ijk}\del_{\m}
\ph^i\del_{\n}\ph^j\x^k+\nabla_iV\x^i), &\ebe a\cr
S^{(2)}&={1\over2}\int d^2x\{g_{ij}D_{\m}\x^iD^{\m}\x^j
+(\g^{\m\n}R_{iklj}+\e^{\m\n}\nabla_lH_{ijk})\del_{\m}
\ph^i\del_{\n}\ph^j\x^k\x^l
\cr&\quad+2\e^{\m\n}H_{ijk}\del_{\m}\ph^iD^{\m}\x^j\x^k+\nabla_i\nabla_jV
\x^i\x^j\},
&\ebe b \cr
S^{(3)}&=\int d^2x({1\over3}\e^{\m\n}H_{ijk}D_{\m}\x^i
D_{\n}\x^j\x^k+{1\over6}\nabla_i\nabla_j\nabla_kV\x^i\x^j\x^k)
+\ldots,  &\ebe c \cr
S^{(4)}&=\int
d^2x({1\over6}\g^{\m\n}R_{iklj}+{1\over4}\e^{\m\n}\nabla_lH_{ijk})
D_{\m}\x^iD_{\n}\x^j\x^k\x^l\cr&\quad
+{1\over{24}}\nabla_i\nabla_j\nabla_k\nabla_l
V\x^i\x^j\x^k\x^l
)+\ldots  ,&\ebe d\cr}$$  }
We omit the subscript $0$ on $\ph^i_0$ here and henceforth. The operator
$D_{\m}$ is defined by
\eqn\ebf{
D_{\m}\x^i=\del_{\m}\x^i-\G^i{}_{jk}\del_{\m}\ph^j\x^k .}
   In Eqs. \ebe{c}\ and \ebe{d}\
we have omitted terms involving $\del_{\m}\ph^i$, since these
are irrelevant for our calculation. Feynman diagrams are constructed using
2-, 3- and 4-point vertices corresponding to the quadratic,
cubic and quartic terms in $S^{(2)}$,
$S^{(3)}$ and $S^{(4)}$ (except for the first term in $S^{(2)}$, which
supplies the propagators used to link the vertices). The 1- and 2-loop
diagrams contributing to the tachyon $\b$-function are shown in Figs. 1
and 2. The corresponding contributions to $\O$ are given by\ref\df{D. Friedan,
\prl45 (1980) 1057;\apny163 (1985) 318.}\ref\aatb{A. A. Tseytlin, \plb178
(1986) 34.}\hugh
\eqn\ebfa{\eqalign{
\O^{(1)}&=-{1\over{\l}}\nabla^2 \cr
\O^{(2)}&={2\over{\l^2}}H^{kmn}H^l{}_{mn}\nabla_k\nabla_l\cr}  }
The 3-loop diagrams
contributing to the tachyon $\b$-function are depicted in Fig. 3. We
subtract subdivergences on a diagram-by-diagram basis. The results for the
divergent contributions $W_{(a)}^{(3)}$--$W_{(g)}^{(3)}$ of the diagrams
Figs. 3(a)--(f) respectively, incorporating the corresponding subtractions,
are:
\eqna\ebg{$$\eqalignno{
W_{(a)}^{(3)}&={4\over3}{1\over{\l^3}}\bigl({1\over{\e^2}}-{5\over4}
{1\over{\e}}\bigr)H^{kmn}H^{lpq}H_{mpr}H_{nq}{}^r\nabla_k\nabla_lV, &\ebg a\cr
W_{(b)}^{(3)}&=-{4\over3}{1\over{\l^3}}\bigl({1\over{\e^3}}-{2\over{\e^2}}
+{7\over4}{1\over{\e}}\bigr)H^{prs}H^q{}_{rs}H_{pm}{}^kH_q{}^{ml}
\nabla_k\nabla_lV, &\ebg b \cr
W_{(c)}^{(3)}&=-{4\over3}{1\over{\l^3}}\bigl({1\over{\e^3}}-{1\over{\e^2}}
\bigr)H^{kmn}H^p{}_{mn}H^{lrs}H_{prs}\nabla_k\nabla_lV &\ebg c \cr
W_{(d)}^{(3)}&=-{2\over3}{1\over{\l^3}}\bigl({1\over{\e^3}}-{1\over{\e^2}}\bigr)
R^{mpqn}H^k{}_{mp}H^l{}_{nq}\nabla_k\nabla_lV &\ebg d \cr
W_{(e)}^{(3)}&=-{4\over3}{1\over{\l^3}}\bigl({1\over{\e^3}}-2{1\over{\e^2}}
+{7\over4}{1\over{\e}}\bigr)R^{kpqm}H^{nl}{}_pH_{nmq}\nabla_k\nabla_lV &\ebg e
 \cr
W_{(f)}^{(3)}&=-{1\over{\l^3}}\bigl({2\over9}{1\over{\e^3}}
-{1\over9}{1\over{\e^2}}+{1\over2}{1\over{\e}}\bigr)R^{kmnp}R^l{}_{mnp}\nabla_k
\nabla_lV &\ebg f\cr
W_{(g)}^{(3)}&={1\over{\l^3}}\bigl({2\over9}{1\over{\e^2}}-{5\over18}
{1\over{\e}}\bigr)
\nabla^kH_{mnp}\nabla^lH^{mnp}\nabla_k\nabla_lV \cr&\quad
+{1\over{\l^3}}\bigl({4\over3}{1\over{\e^3}}-{2\over{\e^2}}+{1\over2}
{1\over{\e}}\bigr)
\nabla^pH^{kmn}\nabla_pH^l{}_{mn}\nabla_k\nabla_lV &\ebg g\cr
W_{(h)}^{(3)}&={1\over{\l^3}}\bigl({4\over3}{1\over{\e^3}}-{8\over3}
{1\over{\e^2}}
+{4\over3}{1\over{\e}}\bigr)\nabla^kH^{lnp}H^m{}_{np}\nabla_{(k}\nabla_l
\nabla_{m)}
&\ebg h \cr
}$$ }
This leads to the following contribution to $\O$ at 3 loops:
\eqn\ebh{\eqalign{
\O^{(3)}&={1\over{\l^3}}\bigl(-{3\over2}R^{kmnp}R^l{}_{mnp}
-7R^{kpqm}H^{nl}{}_pH_{nmq}\cr
&\quad-5H^{kmn}H^{lpq}H_{mpr}H_{nq}{}^r
        -7H^{prs}H^q{}_{rs}H_{pm}{}^kH_q{}^{ml}
       +{3\over2}\nabla^pH^{kmn}\nabla_pH^l{}_{mn}\cr
&\quad        -{5\over6}\nabla^kH_{mnp}\nabla^lH^{mnp}\bigr)
\nabla_k\nabla_l
+{4\over{\l^3}}\nabla^kH^{lnp}H^m{}_{np}\nabla_{(k}\nabla_l\nabla_{m)}  \cr}  }
Specialising to the case of the WZW model, we readily find, using Eqs. \eakb,
 \eakd\ and \eake,
\eqna\ebi{$$\eqalignno{
\O_{WZW(1)}&=-{1\over k}\nabla^2   &\ebi a \cr
\O_{WZW(2)}&={1\over{2k^2}}c^{\cG}\nabla^2 &\ebi b \cr
\O_{WZW(3)}&=-{1\over{4k^3}}c^{\cG2}\nabla^2 &\ebi c \cr }$$  }
which is consistent with the expansion of Eq. \eat\ up to $O({1\over{k^3}})$.
\newsec{Conclusions}
Our central result is the exact expression for the tachyon $\b$-function
in the WZW model, given by Eq. \eau. This quantity played a crucial role
in a recent paper deriving the exact conformally invariant action for the
non-abelian Toda field theory. We checked this result up to 3rd order in
perturbation theory by calculating the tachyon $\b$-function at this order
for a general non-linear $\s$-model, and then specialising to the case of
the WZW model. The result for the general case seems to us to be of interest
since it is the first full, direct computation of a renormalisation group
quantity at three loops for the general non-linear $\s$-model with torsion.
(The three-loop contribution to the dilaton $\b$-function was, however,
calculated indirectly in Refs. \ref\jj{I. Jack and D. R. T. Jones, \plb200
(1988) 453.}, \ref\htb{C. M. Hull and P. K. Townsend, \npb301 (1988) 197.},
\hugh.)

A more stringent check on the validity of Eq. \eat\ would be provided by a
four-loop calculation of $\b^V$. For instance, Eq. \eap\ is trivially satisfied
at three loops since the term involving three derivatives in $\O^{(3)}$ in
Eq. \ebh\ manifestly vanishes upon specialisation to the WZW model. On the
other
hand, at four loops there are possible four-derivative terms in the general
result for $\O^{(4)}$ (see Ref. \ref\jjm{I. Jack, D. R. T. Jones and
N. Mohammedi, \npb334 (1990) 333} for a calculation of $\O^{(4)}$ in the
torsion free case), and these do not obviously vanish immediately upon
specialisation to the WZW case. A four-loop calculation would be formidably
difficult, however.
\vskip 12pt
\line{\bf Acknowledgements}
\vskip 5pt
The Feynman diagrams were produced using the package FeynDiagram written by
Bill Dimm. One of us (I. J.) thanks the S.E.R.C. for financial support.
\listrefs
\line{\bf Figure Captions \hfil}
\item {Fig. 1.} One-loop diagram contributing to tachyon $\b$-function (with
tachyon insertion denoted by cross).
\item {Fig. 2.} Two-loop diagram contributing to tachyon $\b$-function.
\item {Fig. 3.} Three-loop diagrams contributing to tachyon $\b$-function.

\bye